\documentclass[aps,prd,onecolumn,showpacs,showkeys,amsmath,amssymb]{revtex4}
\usepackage{graphicx}
\usepackage{multirow}
\usepackage{dcolumn}
\usepackage{bm}
\usepackage[colorlinks, citecolor=blue,anchorcolor=red,menucolor=red, linkcolor=red,filecolor=red,runcolor=red,urlcolor=blue,frenchlinks=red]{hyperref}

\begin{document}

%=====================================================================================
%=====================================================================================
\title{Chiral corrections to the $1^{-+}$ exotic meson mass}
%=====================================================================================
%=====================================================================================

\author{Bin Zhou}
\email{binzhou@pku.edu.cn} \affiliation{Department of Physics,
Peking University, Beijing 100871, China}
\author{Zhi-Feng Sun}
\email{zhifeng@ific.uv.es} \affiliation{Departamento de F\'{\i}sica
Te\'orica and IFIC, Centro Mixto Universidad de Valencia-CSIC
Institutos de Investigaci\'on de Paterna, Aptdo. 22085, 46071
Valencia, Spain}
\author{Xiang Liu}
\email{xiangliu@lzu.edu.cn} \affiliation{
School of Physical Science and Technology, Lanzhou University, Lanzhou 730000, China\\
Research Center for Hadron and CSR Physics, Lanzhou University and
Institute of Modern Physics of CAS, Lanzhou 730000, China }
\author{Shi-Lin Zhu}
\email{zhusl@pku.edu.cn} \affiliation{Department of Physics and
State Key Laboratory of Nuclear Physics and Technology and
Collaborative Innovation Center of Quantum Matter, Peking
University, Beijing 100871, China}
%\date{\today}

\begin{abstract}

We first construct the effective chiral Lagrangians for the $1^{-
+}$ exotic mesons. With the infrared regularization scheme, we
derive the one-loop infrared singular chiral corrections to the
$\pi_1(1600)$ mass explicitly. We investigate the variation of the
different chiral corrections with the pion mass under two schemes.
Hopefully, the explicit non-analytical chiral structures will be
helpful to the chiral extrapolation of the lattice data from the
dynamical lattice QCD simulation of either the exotic light hybrid
meson or tetraquark state.

\end{abstract}
\maketitle
\pagenumbering{arabic}
%

%%%%%%%%%%%%%%%%%%%%%%%%%%%%%%%%%%
\section{Introduction}\label{sec1}
%%%%%%%%%%%%%%%%%%%%%%%%%%%%%%%%%%

According to the naive non-relativistic quark model, the meson is
composed of a pair of quark and anti-quark. The neutral mesons do
not carry the quantum numbers such as $J^{PC}=0^{--}, 0^{+-}, 1^{-
+}, 2^{- +}...$. In contrast, the non-conventional mesons such as
the hybrid meson, tetraquark states and glueballs are allowed in
quantum chromodynamics (QCD) and can have these quantum numbers.
Sometimes these states are denoted as the exotic states in order to
emphasize the difference from the mesons within the quark model. In
fact, the exotic quantum numbers provide a powerful handle to probe
the non-perturbative behavior of QCD
\cite{Zhu:2007wz,Chen:2016qju,Asner:2008nq}. In this work we focus
on the the exotic meson with $J^{PC}=1^{-+}$, which is a good
candidate of the hybrid meson and tetraquark state.

There are three candidates with $J^{PC}=1^{-+}$: $\pi_1(1400)$,
$\pi_1(1600)$ and $\pi_1(2000)$. Their masses and widths are
($1376\pm17$, $300\pm40$) MeV, ($1653^{\Large +18}_{\Large -15}$,
$225^{\Large +45}_{\Large -28}$) MeV and ($2014 \pm 20 \pm 16$, $230
\pm 21 \pm 73$) MeV~\cite{Amsler:2008zzb} respectively.
$\pi_1(1600)$ was first observed in the reaction
$\pi^{-}p\rightarrow\pi^{-}\pi^{-}\pi^{+}p$ in 1998
\cite{Adams:1998ff,Chung:2002pu}. Later the $\pi_{1}(1600)$ was
confirmed in the $\eta'\pi$ \cite{Ivanov:2001rv}, $f_{1}(1285)\pi$
\cite{Kuhn:2004en,Alekseev:2009aa} and $b_{1}(1235)\pi$
channels\cite{Lu:2004yn,Nozar:2008aa}. Some experiments also
indicated the possible existence of $\pi_1(1400)$
\cite{Chung:1999we,Abele:1999tf,Adams:2006sa} and $\pi_1(2000)$
\cite{Kuhn:2004en}. The existence of $\pi_1(2000)$ awaits further
experimental confirmation. This state was not included in the PDG
since 2010~\cite{Nakamura:2010zzi}.

The current status of the $\pi_1(1400)$ and $\pi_1(1600)$ is a
little murky. There exist speculations that the $\pi_1(1400)$ might
be non-resonant or it may be a tetraquark candidate instead of a
hybrid meson. Although there also exist other possible theoretical
explanations such as a tetraquark candidate
\cite{Chen:2008qw,Chen:2008ne} or a molecule/four-quark mixture
\cite{Narison:2009vj}, the $\pi_1(1600)$ remains a popular candidate
of the light hybrid meson \cite{Meyer:2010ku}. The present calculation
is based on the following three facts: the $1^{-+}$ exotic quantum number,
the SU(3) flavor structure and the current available decay modes.
In other words, it's applicable to all possible interpretations of the $\pi_1$ mesons.

There are many investigations of the $1^{-+}$ light hybrid meson
mass in literature
\cite{Isgur:1984bm,Close:1994hc,Barnes:1995hc,Page:1998gz,Lacock:1998be,McNeile:1998cp,Michael:2003xg,Ebert:2009ub,Kim:2008qh,Ping:2009zza,Kitazoe:1983xx,Dudek:2010wm,Dudek:2009qf,Kisslinger:2009pw}.
The $1^{-+}$ mass extracted from the quenched lattice QCD simulation
ranges from 1.74 GeV~\cite{Hedditch:2005zf} and 1.8
GeV~\cite{Bernard:2003jd} to 2 GeV~\cite{McNeile:1998cp}, which is
significantly larger than the experimental value. This apparent
discrepancy is slightly disturbing. One possible reason may be due
to the fact that all these lattice QCD simulations were performed
with quenched configurations and rather large pion mass on the
lattice. One may wonder whether such a discrepancy may be removed
with dynamical lattice QCD simulations using physical pion
mass. Then one may make chiral extrapolations to extract the
physical mass of the hybrid meson.

In this work we shall derive the explicit expressions of the
non-analytical chiral corrections to the $\pi_1(1600)$ mass up to
one-loop order, which may be used to make the chiral extrapolations
if the dynamical lattice QCD simulations are available.
Throughout our analysis, we focus on the variation of the $\pi_1(1600)$
meson mass with $m_{u,d}$ or $m_\pi$. In the $SU_F(3)$ chiral limit
$m_{u,d,s} \to 0$, $m_{\pi, \eta} \to 0$. The $SU_F(2)$ chiral limit
is adopted where $m_{u,d} \to 0$ and $m_s$ remains finite.
Then, the eta meson mass does not vanish due to the large strange
quark mass.

This paper is organized as follows. We construct the effective
chiral Lagrangians in Sec.~\ref{sec2} and present the formalism in
Sec.~\ref{sec3}. In Sec.~\ref{sec4}, we present the numerical
results and conclude.

%%%%%%%%%%%%%%%%%%%%%%%%%%%%%%%%%%
\section{Lagrangians}\label{sec2}
%%%%%%%%%%%%%%%%%%%%%%%%%%%%%%%%%%

In order to calculate the chiral corrections to the $\pi_1(1600)$
meson mass up to the one loop order, we first construct the
effective chiral Lagrangian \cite{Huang:2010dc,Chen:2010ic}, which
can be expressed as follows
\begin{eqnarray}
\mathcal{L}&=&\mathcal{L}_0+\mathcal{L}_{\rho\pi}+\mathcal{L}_{b_1\pi}+\mathcal{L}_{f_1\pi}+\mathcal{L}_{\eta\pi}+
\mathcal{L}_{\eta'\pi}+\mathcal{L}_{\pi_1\eta}+\mathcal{L}_{\pi_1\eta'}+...,
\end{eqnarray}
where $\mathcal{L}_0$ is the free part
\begin{eqnarray}
\mathcal{L}_0&=&\partial_\mu \vec{\pi}_{1\nu}\cdot \partial^\mu
\vec{\pi}_{1}^\nu - m_0^2 \vec{\pi}_1^\mu\cdot \vec{\pi}_{1\mu}.
\end{eqnarray}

According to the decay modes of $\pi_1(1600)$, we can write down the
interaction terms
\begin{eqnarray}
\label{Lagrangian1}
\mathcal{L}_{\eta\pi}&=&  g_{\eta\pi}\vec{\pi}_1^\mu\cdot\partial_\mu\vec{\pi}\eta,\\
\mathcal{L}_{\eta'\pi}&=&  g_{\eta'\pi}\vec{\pi}_1^\mu\cdot\partial_\mu\vec{\pi}\eta',\\
\mathcal{L}_{\rho\pi}&=&g_{\rho\pi}\epsilon_{\mu\nu\alpha\beta}\vec{\pi}_1^\mu \times
\partial^\alpha \vec{\rho}^\nu \cdot \partial^\beta \vec{\pi},\\
\mathcal{L}_{b_1\pi}&=&g_{b_1\pi}\vec{\pi}_{1\mu}\times
\vec{b}_{1}^\mu \cdot \vec{\pi},\\
%+ \frac{g^{D}_1}{m_{\pi_1}^2} \partial^2 \vec
%\pi_{1\mu} \times \vec b_1^\mu \cdot \vec \pi
% \nonumber \\&& + \frac{g^{D}_2}{m_{\pi_1}} \partial_\nu \vec \pi_{1\mu}\times \vec b_1^\mu \cdot \partial^\nu \vec \pi
% + g^{D}_3 \vec \pi_{1\mu} \times \vec b_{1}^{\mu}\cdot \partial^2 \vec \pi
% \nonumber \\&& + \frac{g^{D}_4}{m_{\pi_1}} \partial^\nu \vec \pi_{1\mu} \times \vec b_{1\nu}\cdot \partial^\mu \vec \pi, \\
\label{Lagrangian2}
\mathcal{L}_{f_1\pi}&=&g_{f_1\pi}\vec{\pi}_{1\mu}\cdot
\vec{\pi}f_{1}^\mu.
%+ \frac{g^{D}_{1'}}{m_{\pi_1}^2} \partial^2 \vec
%\pi_{1\mu} \cdot \vec \pi f_1^\mu
% \nonumber \\&& + \frac{g^{D}_{2'}}{m_{\pi_1}} \partial_\nu \vec \pi_{1\mu} \cdot \partial^\nu \vec \pi f_1^\mu
% + g^{D}_{3'} \vec \pi_{1\mu} \cdot \partial^2\vec \pi f_{1}^{\mu}
% \nonumber \\&& + \frac{g^{D}_{4'}}{m_{\pi_1}} \partial^\nu \vec \pi_{1\mu}\cdot \partial^\mu \vec \pi f_{1\nu}.
\end{eqnarray}

Because of the chiral symmetry and its spontaneous breaking, all the pionic
coupling constants should vanish when either the pion momentum or
its mass goes to zero. The S-wave coupling constants $g_{b_1\pi}$
and $g_{f_1\pi}$ arise from the finite current quark mass
correction. Therefore, these coupling constants are proportional to
$m_\pi^2$,
\begin{eqnarray}
g_{b_1\pi}=g_{b_1\pi}^{*}m_{\pi}^2,~~~~~g_{f_1\pi}=g_{f_1\pi}^{*}m_{\pi}^2.
\end{eqnarray}

\begin{figure} \centering
\includegraphics[width=0.4\linewidth]{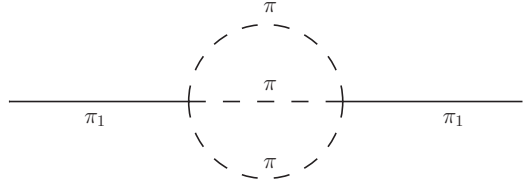}
\caption{The two-loop self energy diagram of the $\pi_1(1600)$ with
three intermediate $\pi$ mesons.} \label{three_pi_loop}
\end{figure}

The $\pi_1\rightarrow\pi\pi\pi$ decay mode may lead to the two-loop
self energy diagram $\pi_1(1600)$ in Fig. \ref{three_pi_loop}. We
ignore the contribution from this diagram since we focus on the
chiral corrections to the $\pi_1(1600)$ mass up to the one-loop
order in this work. Moreover, some contribution of this two-loop
diagram may have been partly included in the one-loop diagram with
the intermediate $\rho$ and $\pi$ meson because the $\rho$ meson is
the two pion resonance.

Further more, we need the chiral interaction between the
$\pi_1(1600)$ and the pseudo scalar mesons, which is similar to the
chiral Lagrangians of the vector mesons
\cite{Klingl:1996by,Ecker:1989yg,Birse:1996hd,Bijnens:1997ni,Meissner:1987ge}.
It should be stressed that the $\pi_1\pi_1\pi$ interaction is
forbidden by the G-parity conservation. We have
\begin{eqnarray}
\mathcal{L}_{\pi_1\eta}&=&g_{\pi_1\eta}\epsilon_{\mu\nu\alpha\beta}\vec{\pi}_1^\mu\cdot
\partial^\alpha\vec{\pi}_1^\nu\partial^\beta\eta,\\
\mathcal{L}_{\pi_1\eta'}&=&g_{\pi_1\eta'}\epsilon_{\mu\nu\alpha\beta}\vec{\pi}_1^\mu\cdot
\partial^\alpha\vec{\pi}_1^\nu\partial^\beta\eta'.
\end{eqnarray}

For the $\pi_1\pi_1\pi\pi$ and $\pi_1\pi_1\eta\eta$ interaction, we
have
\begin{eqnarray}
\mathcal{L}_{\pi_1\pi_1\pi\pi}  &=&
c_{1}m_{\pi}^2\vec{\pi}\cdot\vec{\pi}\vec{\pi}_{1}^{\mu}\cdot\vec{\pi}_{1\mu}
+c_{2}\partial_{\mu}\vec{\pi}\cdot\partial^{\mu}\vec{\pi}\vec{\pi}_{1}^{\nu}\cdot\vec{\pi}_{1\nu}\nonumber~~\\&&
+c_{3}\partial_{\mu}\vec{\pi}\cdot\partial_{\nu}\vec{\pi}\vec{\pi}_{1}^{\mu}\cdot\vec{\pi}_{1}^{\nu}
+\frac{c_{4}}{m_{\pi_1}}\partial_{\mu}\vec{\pi}\cdot\vec{\pi}\vec{\pi}_{1}^{\nu}\cdot\partial^{\mu}\vec{\pi}_{1\nu}\nonumber~~\\&&
+\frac{c_{5}}{m_{\pi_1}}\partial_{\mu}\vec{\pi}\cdot\vec{\pi}\vec{\pi}_{1}^{\nu}\cdot\partial_{\nu}\vec{\pi}_{1}^{\mu}
+\frac{c_{6}m_{\pi}^2}{m_{\pi_1}^2}\vec{\pi}\cdot\vec{\pi}\partial_{\mu}\vec{\pi}_{1}^{\nu}\cdot\partial^{\mu}\vec{\pi}_{1\nu},\\
\mathcal{L}_{\pi_1\pi_1\eta\eta} &=&
c_{1}^{*}m_{\eta}^2\eta^2\vec{\pi}_{1}^{\mu}\cdot\vec{\pi}_{1\mu}
+c_{2}^{*}\partial_{\mu}\eta\partial^{\mu}\eta\vec{\pi}_{1}^{\nu}\cdot\vec{\pi}_{1\nu}\nonumber~~\\&&
+c_{3}^{*}\partial_{\mu}\eta\partial_{\nu}\eta\vec{\pi}_{1}^{\mu}\cdot\vec{\pi}_{1}^{\nu}
+\frac{c_{4}^{*}}{m_{\pi_1}}\partial_{\mu}\eta\eta\vec{\pi}_{1}^{\nu}\cdot\partial^{\mu}\vec{\pi}_{1\nu}\nonumber~~\\&&
+\frac{c_{5}^{*}}{m_{\pi_1}}\partial_{\mu}\eta\eta\vec{\pi}_{1}^{\nu}\cdot\partial_{\nu}\vec{\pi}_{1}^{\mu}
+\frac{c_{6}^{*}m_{\eta}^2}{m_{\pi_1}^2}\eta^2\partial_{\mu}\vec{\pi}_{1}^{\nu}\cdot\partial^{\mu}\vec{\pi}_{1\nu}.
\end{eqnarray}

In order to absorb the divergence in the one-loop chiral
corrections, we need the following counter terms
\begin{eqnarray}
\label{counter} \mathcal{L}_{counter} &=&
e_{1}(m_\pi^2+m_\eta^2)\vec{\pi}_{1\mu}\cdot\vec{\pi}_{1}^{\mu}+
e_{2}(m_\pi^2+m_\eta^2)^2\vec{\pi}_{1\mu}\cdot\vec{\pi}_{1}^{\mu}.
\end{eqnarray}
$\mathcal{L}_{counter}$ is similar to the chiral Lagrangians of the
vector mesons in the form of $\langle\chi_{+}\rangle\langle V_\mu
V^\mu \rangle$ and ${\langle\chi_{+}\rangle}^2\langle V_\mu V^\mu
\rangle$, where $V_\mu$ is the vector meson and the notation
$\chi_{+}$ is related to the current quark mass.

%%%%%%%%%%%%%%%%%%%%%%%%%%%%%%%%%%%%%%%%%%%%%%%%%%%%%%%%%%%%%%%%%%%%
\section{Chiral corrections to the $\pi_1(1600)$ mass}\label{sec3}
%%%%%%%%%%%%%%%%%%%%%%%%%%%%%%%%%%%%%%%%%%%%%%%%%%%%%%%%%%%%%%%%%%%%

With the above preparation, we start to calculate the chiral
corrections to the mass of $\pi_1(1600)$. The propagator of the
$\pi_1(1600)$ is defined as
\begin{eqnarray}
S_0^{\mu\nu}=i\int d^4xe^{ip\cdot x}\langle
0|T\{\pi_1^\mu(x)\pi_1^\nu(0)\}|0 \rangle,
\end{eqnarray}
where p is the four momenta of $\pi_1$. At the lowest order, the
propagator simply reads
\begin{eqnarray}
S_0^{\mu\nu}=\frac{-i(g^{\mu\nu}-p^\mu
p^\nu/m_0^2)}{p^2-m_0^2+i\epsilon}=\frac{-i(g_{\mu\nu}-\frac{p_\mu
p_\nu}{p^2})}{p^2-m_0^2}+\frac{ip_\mu p_\nu}{p^2m_0^2}
\end{eqnarray}
and its inverse is
\begin{eqnarray}
(S_0^{-1})^{\mu\nu}=i((p^2-m_0^2)g^{\mu\nu}-p^\mu p^\nu).
\end{eqnarray}
Here, $m_0$ denotes the bare mass of $\pi_1(1600)$.

We separate the self energy $\Sigma_{\mu\nu}(p^2)$ into the
transversal and longitudinal parts
\begin{eqnarray}
\Sigma_{\mu\nu}(p^2)=\left(g_{\mu\nu}-\frac{p_\mu
p_\nu}{p^2}\right)\Sigma_T(p^2)+\frac{p_\mu
p_\nu}{p^2}\Sigma_L(p^2).
\end{eqnarray}

\begin{figure}
\centering
\includegraphics[width=0.6\linewidth]{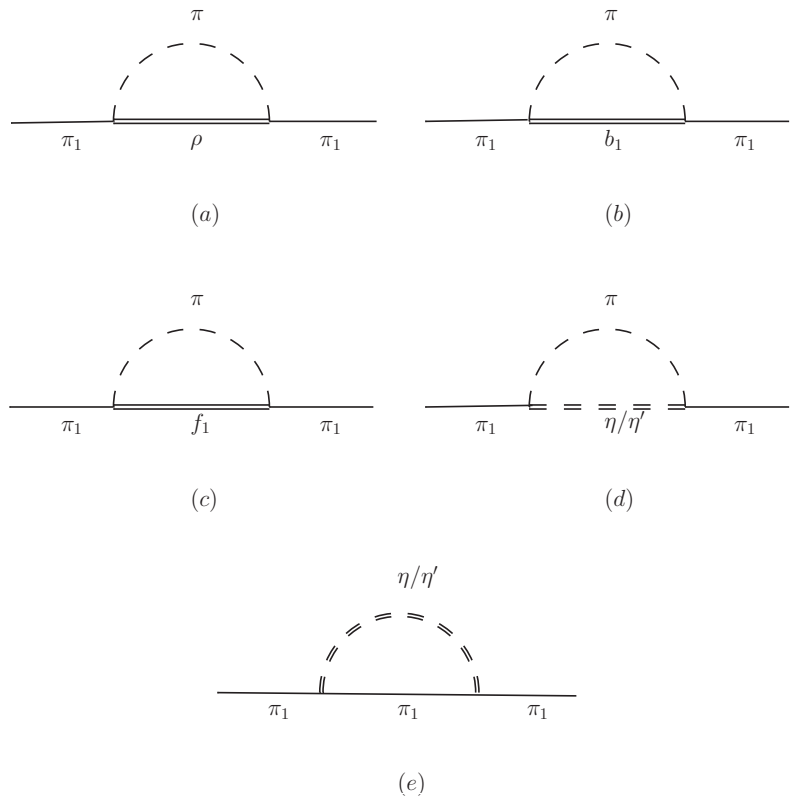}
\caption{The one-loop self energy diagrams of the $\pi_1(1600)$ with
one light meson plus one $\pi$ or $\eta$.} \label{pi1selfenergy}
\end{figure}

The full propagator reads
\begin{eqnarray}
S^{\mu\nu}=S_0^{\mu\nu}+S_0^{\mu\alpha}(i\Sigma)_{\alpha\beta}(p^2)S_0^{\beta\nu}+...=\left[(S_0^{-1}-i\Sigma)^{\mu\nu}\right]^{-1}.
\end{eqnarray}
which can be expressed as
\begin{eqnarray}
S_{\mu\nu}=\frac{-i(g_{\mu\nu}-p_\mu
p_\nu/p^2)}{p^2-m_0^2-\Sigma_T(p^2)}+\frac{ip_\mu
p_\nu}{p^2(m_0^2+\Sigma_L(p^2))}.
\end{eqnarray}
Only the transverse part $\Sigma_T(p^2)$ will shift the pole
position. Therefore we concentrate on the transversal part of the
self energy \cite{Bruns:2013tja} and consider all the Feynman
diagrams shown in Fig. \ref{pi1selfenergy} and Fig.
\ref{counterterms}. The $\pi_1(1600)$ mass satisfies the relation
\begin{eqnarray}
\label{delta} m_{\pi_1}^2-m_0^2-\Sigma_T(m_{\pi_1}^2)=0.
\end{eqnarray}

In order to obtain the quark mass ($\sim m_\pi^2,~m_\eta^2$)
dependence of the self energy corrections, it is convenient to adopt
the infrared regularization (IR) scheme
\cite{Borasoy:2006fk,Bruns:2004tj,Becher:1999he} to calculate the
loop integrals. Usually, the IR method is used in order not to break
the power counting while dealing with the integral. Unfortunately,
there doesn't exist a proper power counting rule for the issue we
are dealing with. There are a few different mass scales such as the
$\pi_1$ mass, the $\pi,~\eta$ meson masses, the masses of other
meson resonances, and the chiral symmetry breaking scale. The mass
of the $\pi_1$ is so high that the $\pi,~\eta$ and other light
mesons can take large momenta, and thus the convergence of a chiral
expansion is not ensured. However, for our purpose, the IR method
still can be used to derive the non-analytic part of an integral.
The non-analytical chiral corrections to the self-energy of the
$\pi_1$ are inherent and intrinsic due to the presence of the chiral
fields, and the non-analytical chiral structures are universal and
model independent to a large extent. One may derive them using very
different theoretical approaches such chiral quark model, effective
chiral Lagrangians at the hadronic level or rigorous chiral
perturbation theory (ChPT). With ChPT, one can include both
analytical and non-analytical corrections order by order with
consistent power counting. In contrast, with the effective chiral
Lagrangians at the hadronic level as employed in this work, there
does not exist consistent power counting. Fortunately, the
non-analytical corrections from different approaches are similar if
one considers the one-loop diagrams. The non-analytical structures
may play an important role in the chiral extrapolation of the
dynamical lattice QCD simulation of the $1^{-+}$ exotic meson mass,
which is sensitive to the pion mass on the lattice. Within the IR
scheme, the so-called 'infrared singular part' turns out to be the
main contribution of the loop integral in the chiral limit. However,
one can also find the full expressions of these loop integrals by
performing the standard Lorentz invariant calculation in
Refs.\cite{Bernard:1995dp,Fuchs:2003qc}.

For a certain diagram, there are three mass scales, $M_{\pi_1}$ and
the masses of the two intermediate states $m, M$. We assume $M> m$.
The main contribution of a loop integral comes from the poles of the
propagators, which are called as the 'soft poles' and 'hard poles'
in Refs. \cite{Tang:1996ca,Ellis:1997kc}.

When one expands the loop integral in terms of the small parameters
such as $m/M$ or $m/\mu$ where $\mu$ is the renormalization scale,
one notices that the 'soft part' contribution contains all the terms
which are non-analytic in the expansion parameter. In contrast,
the 'hard part' is a local polynomial in these parameters which can be
absorbed by the low energy constants of higher order Lagrangians
\cite{Becher:1999he}.

Since we are interested in the small chiral fluctuations around the
mass shell of $\pi_1(1600)$, we set the kinematical region $p^2\sim
M_{\pi_1}^2$. In particular, we set the the regularization scale to
be $M_{\pi_1}$. These self-energy diagrams can be divided into two
categories. The first class of diagrams fulfills the condition
$M_{\pi_1}^2\gg (M+m)^2$ and $m^2\ll M^2$, including those diagrams
with the $\rho\pi, \eta\pi, b_1(1235)\pi, f_1(1285)\pi$ and
$\eta'\pi$ as the intermediate states. The second class corresponds
to the condition $M_{\pi_1}^2\sim M^2$ and $ m^2\ll M^2 $, where the
intermediate states are the $\pi_1(1600)\eta$ and $\pi_1(1600)
\eta'$.

\subsection{The light meson pion loop}

Now we deal with the light meson pion loop integration corresponding
to diagrams (a)-(d) in Fig. \ref{pi1selfenergy}. Consider the scalar
loop integrals
\begin{eqnarray}
I_{\pi X}(p^2)&=&\mu^{4-d}\int
\frac{d^dl}{(2\pi)^d}\frac{1}{[l^2-m_\pi^2+i\epsilon][(p-l)^2-M^2+i\epsilon]},
\end{eqnarray}
where X represents the $\rho, b_1, f_1, \eta'$ mesons. $l$ and
$p$ denote the loop momentum and external momentum respectively.
After performing the $l$-integration, the above integral reads
\begin{eqnarray}
I_{\pi X}(p^2)&=& \mu^{4-d}\Gamma \left(2-\frac{d}{2}\right)\frac{i
M^{d-4}}{(4\pi)^{\frac{d}{2}}}\int_{0}^{1}dx
(\Delta)^{\frac{d}{2}-2}
\end{eqnarray}
with
\begin{eqnarray}
\Delta &=& b x^2-(a+b-1)x+a,\nonumber\\
a&=& \frac{m_\pi^{2}}{M^{2}}~,~ b = \frac{p^{2}}{M^{2}} .
\end{eqnarray}
Since we choose the external momentum $p$ near the mass shell of
$\pi_1(1600)$, we always have $(p^2-m_\pi^2+M^2)^2-4p^2M^2>0$.
$\Delta$ can be re-expressed as $\Delta=b(x-x_1)(x-x_2)$, with
\begin{equation}
x_{1,2} =  \frac{a + b -1}{2 b} \left( 1 \pm
         \sqrt{1 - \frac{4 a b}{(a+ b -1)^2} }  \  \right)~.
\end{equation}
Obviously we have $0< x_2 < x_1 <1$. We now divide the integral into
three parts according to the integration interval
\begin{eqnarray}
I_{\pi X} = \mu^{4-d}\Gamma \left(2-\frac{d}{2}\right)\frac{i
M^{d-4}}{(4\pi)^{\frac{d}{2}}}
           \left( I_{\pi X}^{(1)} + I_{\pi X}^{(2)} + I_{\pi X}^{(3)}  \right)
\end{eqnarray}
with
\begin{eqnarray}
I_{\pi X}^{(1)} (p^2)  &=& \int_0^{x_2} dx \ \left[ b (x - x_1) (x -
x_2)
                                   \right]^{\frac{d}{2}-2}~, \nonumber \\
I_{\pi X}^{(2)} (p^2)  &=& \int_{x_2}^{x_1} dx \ \left[ b (x - x_1)
(x - x_2)
                                       \right]^{\frac{d}{2}-2}~, \nonumber \\
I_{\pi X}^{(3)} (p^2)  &=& \int_{x_1}^1 dx \ \left[ b (x - x_1) (x -
x_2) \right]^{\frac{d}{2}-2} ~.
\end{eqnarray}

We first consider $I_{\pi X}^{(1)}$. The assumption $p^2\gg
(M+m_\pi)^2$ and $m_\pi^2\ll M^2$ leads to
\begin{equation}
a\ll 1, ~~~~ \frac{4 a b}{(a+ b -1)^2} \ll 1.
\end{equation}
So we can expand $x_{1,2}$ in terms of the small parameter $a$,
\begin{eqnarray}
x_1 &=& \frac{b -1}{b} - \frac{a}{b(b -1)} - \frac{a^2}{(b -1)^3}
         + {\cal O}(a^3) \ , \nonumber \\
x_2 &=& \frac{a}{b -1} + \frac{a^2}{(b -1)^3}
         + {\cal O}(a^3) \ .
\end{eqnarray}
Then we have
\begin{eqnarray}
I_{\pi X}^{(1)}(p^{2}) & = &
(-bx_{1})^{\frac{d}{2}-2}\intop_{0}^{x_{2}}dx[(1-x/x_{1})(x-x_{2})]^{\frac{d}{2}-2}.
\end{eqnarray}
Recall that $x_1 \sim {\cal O}(1)$ and $x_2 \sim {\cal O}(a)$. When
$x \in [0,x_2]$, we can expand the above integral in terms of the
parameter $x/x_1$
\begin{eqnarray}
I_{\pi X}^{(1)}(p^{2}) & = &
(-bx_{1})^{\frac{d}{2}-2}\intop_{0}^{x_{2}}dx(x-x_{2})^{\frac{d}{2}-2}\sum_{m=0}^{\infty}\frac{\Gamma(\frac{d}{2}-1)}{\Gamma(\frac{d}{2}-1-m)m!}(-\frac{x}{x_{1}})^{m}.
\end{eqnarray}
After the interchange of summation and integration, we get
\begin{eqnarray}
I_{\pi X}^{(1)}(p^{2}) =  (bx_{1})^{\frac{d}{2}-1}x_{2}^{\frac{d}{2}-1}\sum_{m=0}^{\infty}\frac{\Gamma(\frac{d}{2}-1)\Gamma(\frac{d}{2}-1)}{\Gamma(\frac{d}{2}-1-m)\Gamma(\frac{d}{2}+m)}(-\frac{x_{2}}{x_{1}})^{m}.
\end{eqnarray}
Clearly $I_{\pi X}^{(1)}$ is non-analytic in $a$ for noninteger dimension $d$.

We move on to the $I_{\pi X}^{(2)}$ part. After shifting the
integration variable, we get
\begin{eqnarray}
I_{\pi X}^{(2)}(p^2) &
= &
(-b)^{\frac{d}{2}-2}\intop_{0}^{x_{1}-x_{2}}dx[x(x_{1}-x_{2}-x)]^{\frac{d}{2}-2}.
\end{eqnarray}
With the replacement $x=(x_1 - x_2)y$, one gets
\begin{eqnarray}
I_{\pi X}^{(2)}(p^2)& = & (-b)^{\frac{d}{2}-2} (x_{1}-x_{2})^{d-3} \intop_{0}^{1}dy[y(1-y)]^{\frac{d}{2}-2}~\nonumber\\
       & = & (-b)^{\frac{d}{2}-2}(x_{1}-x_{2})^{d-3}\frac{[\Gamma(\frac{d}{2}-1)]^{2}}{\Gamma(d-2)}.
\end{eqnarray}
$I_{\pi X}^{(2)}$ is complex and proportional to $(x_{1}-x_{2})^{d-3}$ that can
be expanded in powers of $x_2$.

We expand the third integral $I_{\pi X}^{(3)}$ in terms of $x_2 /
x$, i.e.,
\begin{eqnarray}
I_{\pi X}^{(3)}(p^{2}) & = & \intop_{x_{1}}^{1}dx[b(x-x_{1})]^{\frac{d}{2}-2}x^{\frac{d}{2}-2}(1-\frac{x_{2}}{x})^{\frac{d}{2}-2}~\nonumber\\
& = & \intop_{x_{1}}^{1}dx[b(x-x_{1})]^{\frac{d}{2}-2}x^{\frac{d}{2}-2}\sum_{m=0}^{\infty}\frac{\Gamma(\frac{d}{2}-1)}{\Gamma(\frac{d}{2}-1-m)m!}(\frac{x_{2}}{x})^{m}~\nonumber\\
 & = & \sum_{m=0}^{\infty}\frac{\Gamma(\frac{d}{2}-1)}{\Gamma(\frac{d}{2}-1-m)m!}x_{2}^{m}\intop_{x_{1}}^{1}dx[b(x-x_{1})]^{\frac{d}{2}-2}x^{\frac{d}{2}-2-m}.
\end{eqnarray}
Obviously $I_{\pi X}^{(3)}$ only contains the integer powers of $a$.

It is clear that $I_{\pi X}^{(3)}$ and the real part of $I_{\pi X}^{(2)}$ are regular in $a$ and will
not produce any infrared singular terms for an arbitrary value of the dimension $d$. Thus these parts
can be absorbed into the low energy constants of the effective Lagrangian. On the other hand,
$I_{\pi X}^{(1)}$ develops an infrared singularity as $a\rightarrow 0$ for negative enough dimension d.
This part is the so-called 'infrared singular part' of $I_{\pi X}$ in the IR method of Refs. \cite{Borasoy:2006fk,Bruns:2004tj,Becher:1999he}.
The 'infrared singular part' contains all the terms which are non-analytic in $a$ as the typical chiral log terms $\ln a$,
such terms can not be absorbed into the low energy constants of the effective Lagrangian. Furthermore,
the contribution of the 'infrared singular part' dominates the $I_{\pi X}$ as $a\rightarrow 0$.

Finally we obtain the 'infrared singular part' in $I_{\pi X}$ with the imaginary part,
\begin{eqnarray}
I_{\pi X}^{IR}(p^{2}) & = & \frac{i}{16\pi^{2}}x_{2}\left[L+1-\ln(\frac{m_{\pi}^{2}}{\mu^{2}})+(\frac{x_{1}-x_{2}}{x_{2}})\ln(\frac{x_{1}-x_{2}}{x_{1}})\right]-\frac{1}{16\pi}(x_1-x_2)~\nonumber \\
& = & \frac{i}{16\pi^{2}}x_{2}\left[L-\ln(\frac{m_{\pi}^{2}}{\mu^{2}})\right]+\frac{i}{16\pi^{2}}\left[x_{2}-(x_{1}-x_{2})(\frac{x_{2}}{x_{1}}+\frac{1}{2}\frac{x_{2}^{2}}{x_{1}^{2}})\right]-\frac{1}{16\pi}(x_1-x_2)~ \nonumber \\
 & = & \frac{i}{16\pi^{2}}\left[L-\ln(\frac{m_{\pi}^{2}}{m_{\pi_1}^{2}})\right](\frac{a}{b-1}+\frac{a^{2}}{(b-1)^{3}})+\frac{i}{32\pi^{2}}\frac{a^{2}b}{(b-1)^{3}}
 \nonumber \\&&-\frac{1}{16\pi}\left[\frac{b-1}{b}-\frac{(b+1)a}{b(b-1)}-\frac{2a^2}{(b-1)^3}\right]+O(a^{3}),
\end{eqnarray}
where $ L  = \frac{1}{\epsilon}-\gamma_{E}+\ln4\pi+1$ and we let
$\mu =m_{\pi_1}$.

Up to $\cal O$($m_\pi^4$) and $\cal O$($m_\eta^4$), we collect the
one-loop chiral corrections to the self-energy of the
$\pi_1(1600)$ below
\begin{eqnarray}
\Sigma_{T,IR}^{\rho\pi}(m_{\pi_{1}}^{2})&=&\frac{g_{\rho\pi}^{2}m_{\pi_{1}}^{2}m_{\pi}^{4}}{32\pi^{2}(m_{\pi_{1}}^{2}-m_{\rho}^{2})}\left[1-2\ln(\frac{m_{\pi}^{2}}{m_{\pi_{1}}^{2}})\right]~\nonumber \\ &&
-i g_{\rho\pi}^2\left[\frac{(m_{\pi_1}^2-m_{\rho}^2)^3}{48\pi m_{\pi_1}^2}-\frac{m_{\pi}^2(m_{\pi_1}^4-m_{\rho}^4)}{16\pi m_{\pi_1}^2}
+\frac{m_{\pi}^4(m_{\pi_1}^4+m_{\rho}^4)}{16\pi m_{\pi_1}^2(m_{\pi_1}^2-m_{\rho}^2)}\right],
\end{eqnarray}
\begin{eqnarray}
\Sigma_{T,IR}^{\pi\eta'}(m_{\pi_{1}}^{2}) &=&
\frac{g_{\eta'\pi}^{2}m_{\pi}^{4}}{128\pi^{2}(m_{\pi_{1}}^{2}-m_{\eta'}^{2})}\left[1-2\ln(\frac{m_{\pi}^{2}}{m_{\pi_{1}}^{2}})\right]~\nonumber \\ &&
-i g_{\eta'\pi}^2\left[\frac{(m_{\pi_1}^2-m_{\eta'}^2)^3}{192\pi m_{\pi_1}^4}-\frac{m_{\pi}^2(m_{\pi_1}^4-m_{\eta'}^4)}{64\pi m_{\pi_1}^4}
+\frac{m_{\pi}^4(m_{\pi_1}^4+m_{\eta'}^4)}{64\pi m_{\pi_1}^4(m_{\pi_1}^2-m_{\eta'}^2)}\right],
\end{eqnarray}
\begin{eqnarray}
\Sigma_{T,IR}^{b_{1}\pi}(m_{\pi_{1}}^{2}) &=&
g_{b_{1}\pi}^{2}\left\{\frac{m_{\pi}^{4}(m_{\pi_{1}}^{4}-6m_{\pi_{1}}^{2}m_{b_{1}}^{2}+m_{b_{1}}^{4})}{64\pi^{2}m_{b_{1}}^{2}(m_{\pi_{1}}^{2}-m_{b_{1}}^{2})^{3}}\right.
~\nonumber \\ && +\left.\left[\frac{m_{\pi}^{2}}{8\pi^{2}(m_{\pi_{1}}^{2}-m_{b_{1}}^{2})}-\frac{m_{\pi}^{4}(m_{\pi_{1}}^{4}-2m_{\pi_{1}}^{2}m_{b_{1}}^{2}-3m_{b_{1}}^{4})}{32\pi^{2}m_{b_{1}}^{2}(m_{\pi_{1}}^{2}-m_{b_{1}}^{2})^{3}}\right]\ln(\frac{m_{\pi}^{2}}{m_{\pi_{1}}^{2}})\right\} ~\nonumber \\ &&
-i g_{b_{1}\pi}^2\left[\frac{(m_{\pi_1}^2-m_{b_1}^2)(m_{\pi_1}^4+10m_{\pi_1}^2m_{b_1}^2+m_{b_1}^4)}{96\pi m_{b_1}^2 m_{\pi_1}^4}\right.
~\nonumber \\ &&
-\left.\frac{m_{\pi}^2(m_{\pi_1}^2+m_{b_1}^2)^3}{32\pi m_{b_1}^2 m_{\pi_1}^4 (m_{\pi_1}^2-m_{b_1}^2)}+\frac{m_{\pi}^4(m_{\pi_1}^2+m_{b_1}^2)^2(m_{\pi_1}^4-4 m_{\pi_1}^2 m_{b_1}^2+m_{b_1}^4)}{32\pi m_{b_1}^2 m_{\pi_1}^4 (m_{\pi_1}^2-m_{b_1}^2)^3}\right],
\end{eqnarray}
\begin{eqnarray}
\Sigma_{T,IR}^{f_{1}\pi}(m_{\pi_{1}}^{2}) &=&
g_{f_{1}\pi}^{2}\left\{\frac{m_{\pi}^{4}(m_{\pi_{1}}^{4}-6m_{\pi_{1}}^{2}m_{f_{1}}^{2}+m_{f_{1}}^{4})}{128\pi^{2}m_{f_{1}}^{2}(m_{\pi_{1}}^{2}-m_{f_{1}}^{2})^{3}}\right.
~\nonumber \\ &&+\left.\left[\frac{m_{\pi}^{2}}{16\pi^{2}(m_{\pi_{1}}^{2}-m_{f_{1}}^{2})}-\frac{m_{\pi}^{4}(m_{\pi_{1}}^{4}-2m_{\pi_{1}}^{2}m_{f_{1}}^{2}-3m_{f_{1}}^{4})}{64\pi^{2}m_{f_{1}}^{2}(m_{\pi_{1}}^{2}-m_{f_{1}}^{2})^{3}}\right]\ln(\frac{m_{\pi}^{2}}{m_{\pi_{1}}^{2}})\right\}
~\nonumber \\ &&
-i g_{f_{1}\pi}^2\left[\frac{(m_{\pi_1}^2-m_{f_1}^2)(m_{\pi_1}^4+10m_{\pi_1}^2m_{f_1}^2+m_{f_1}^4)}{192\pi m_{f_1}^2 m_{\pi_1}^4}\right.
~\nonumber \\ &&
-\left.\frac{m_{\pi}^2(m_{\pi_1}^2+m_{f_1}^2)^3}{64\pi m_{f_1}^2 m_{\pi_1}^4 (m_{\pi_1}^2-m_{f_1}^2)}+\frac{m_{\pi}^4(m_{\pi_1}^2+m_{f_1}^2)^2(m_{\pi_1}^4-4 m_{\pi_1}^2 m_{f_1}^2+m_{f_1}^4)}{64\pi m_{f_1}^2 m_{\pi_1}^4 (m_{\pi_1}^2-m_{f_1}^2)^3}\right].
\end{eqnarray}

\subsection{$\eta$-$\pi$ loop}
Consider the scalar loop integral for $\eta$-$\pi$ loop
\begin{eqnarray}
I_{\pi \eta}(p^2)=&=&\mu^{4-d}\int
\frac{d^dl}{(2\pi)^d}\frac{1}{[l^2-m_\pi^2+i\epsilon][(p-l)^2-m_{\eta}^2+i\epsilon]},
\end{eqnarray}
After performing the l-integration, the above integral reads
\begin{eqnarray}
I_{\pi \eta}(p^2)&=& \mu^{4-d}\Gamma \left(2-\frac{d}{2}\right)\frac{i
p^{d-4}}{(4\pi)^{\frac{d}{2}}}\int_{0}^{1}dx
(\Delta)^{\frac{d}{2}-2}
\end{eqnarray}
with
\begin{eqnarray}
\Delta &=& x^2-(a-b+1)x+a,\nonumber\\
a&=& \frac{m_\pi^{2}}{p^{2}}~,~ b = \frac{m_\eta^{2}}{p^{2}} .
\end{eqnarray}
Similarly, $\Delta$ can be re-expressed as $\Delta=b(x-x_1)(x-x_2)$, with
\begin{eqnarray}
x_{1,2}=\frac{a-b+1}{2}\left(1 \pm \sqrt{1-\frac{4a}{(a-b+1)^2}}\right)
\end{eqnarray}
Obviously we have
\begin{eqnarray}
a\ll 1,~~~b\ll 1,~~~\frac{4a}{(a-b+1)^2}\ll 1.
\end{eqnarray}
So we can expand $x_{1,2}$ in terms of $a$ and $b$
\begin{eqnarray}
x_1 &=& 1-b-ab+\ldots, \nonumber\\
x_2 &=& a+ab+\ldots.
\end{eqnarray}
With the same method, we divide the integral into three parts
\begin{eqnarray}
I_{\pi \eta} = \mu^{4-d}\Gamma \left(2-\frac{d}{2}\right)\frac{i
p^{d-4}}{(4\pi)^{\frac{d}{2}}}
           \left( I_{\pi \eta}^{(1)} + I_{\pi \eta}^{(2)} + I_{\pi \eta}^{(3)}  \right)
\end{eqnarray}
with
\begin{eqnarray}
I_{\pi \eta}^{(1)} (p^2)  &=& \int_0^{x_2} dx \ \left[ (x - x_1) (x -
x_2)\right]^{\frac{d}{2}-2}~\nonumber \\
&=& x_{1}^{\frac{d}{2}-1}x_{2}^{\frac{d}{2}-1}\sum_{m=0}^{\infty}\frac{\Gamma(\frac{d}{2}-1)
\Gamma(\frac{d}{2}-1)}{\Gamma(\frac{d}{2}-1-m)\Gamma(\frac{d}{2}+m)}(-\frac{x_{2}}{x_{1}})^{m},\\
I_{\pi \eta}^{(2)} (p^2)  &=& \int_{x_2}^{x_1} dx \ \left[ (x - x_1)
(x - x_2)\right]^{\frac{d}{2}-2}~\nonumber \\
&=& (-1)^{\frac{d}{2}-2}(x_{1}-x_{2})^{d-3}\frac{[\Gamma(\frac{d}{2}-1)]^{2}}{\Gamma(d-2)},\\
I_{\pi \eta}^{(3)} (p^2)  &=& \int_{x_1}^1 dx \ \left[ (x - x_1) (x -
x_2) \right]^{\frac{d}{2}-2} ~.
\end{eqnarray}
The $I_{\pi \eta}^{(1)}$ and $I_{\pi \eta}^{(2)}$ are similar for
the case in the previous section, where $I_{\pi \eta}^{(1)}$ belongs
to the 'infrared singular part' of $I_{\pi \eta}$ and $I_{\pi
\eta}^{(2)}$ contains an imaginary part. However, the $I_{\pi
\eta}^{(3)}$ is quite different. To calculate the $I_{\pi
\eta}^{(3)}$, we first shift the integration variable
\begin{eqnarray}
I_{\pi \eta}^{(3)}(p^2) &=& \intop_{0}^{1-x_1}dy\left[(1-x_1-y)(1-x_2-y)\right]^{\frac{d}{2}-2} ~\nonumber\\
&=& (1-x_2)^{\frac{d}{2}-2}\intop_{0}^{1-x_1}dy\left[(1-x_1-y)(1-\frac{y}{1-x_2})\right]^{\frac{d}{2}-2}.
\end{eqnarray}
Since $(1-x_1) \sim {\cal O}(a)\sim {\cal O}(b)$ and $(1-x_2) \sim {\cal O}(1)$. When
$y \in [0,1-x_1]$, we can expand the above integral in terms of the
parameter $y/(1-x_2)$
\begin{eqnarray}
I_{\pi \eta}^{(3)}(p^2) &=& (1-x_2)^{\frac{d}{2}-2}\intop_{0}^{1-x_1}dy(1-x_1-y)^{\frac{d}{2}-2}\sum_{m=0}^{\infty}\frac{\Gamma(\frac{d}{2}-1)}{\Gamma(\frac{d}{2}-1-m)m!}(-\frac{y}{1-x_2})^{m} ~\nonumber \\
&=&(1-x_{1})^{\frac{d}{2}-1}(1-x_{2})^{\frac{d}{2}-1}\sum_{m=0}^{\infty}\frac{\Gamma(\frac{d}{2}-1)\Gamma(\frac{d}{2}-1)}{\Gamma(\frac{d}{2}-1-m)\Gamma(\frac{d}{2}+m)}(-\frac{1-x_{1}}{1-x_{2}})^{m}.
\end{eqnarray}
Obviously $I_{\pi X}^{(3)}$ is non-analytic in $b$ for for
non-integer dimension $d$. In other words, $I_{\pi X}^{(3)}$ also
contributes to the 'infrared singular part'. The 'infrared singular
part' of $I_{\pi \eta}$ with the imaginary part are thus
\begin{eqnarray}
I_{\pi \eta}^{IR}(p^2) &=& \mu^{4-d}\Gamma \left(2-\frac{d}{2}\right)\frac{ip^{d-4}}{(4\pi)^{\frac{d}{2}}}
\left( I_{\pi \eta}^{(1)} + \mbox{Im}(I_{\pi \eta}^{(2)}) + I_{\pi \eta}^{(3)}  \right) ~\nonumber \\
&=& \frac{i}{16\pi^2}(1-x_1)\left[L+1-\ln(\frac{m_{\eta}^{2}}{\mu^2})+\frac{x_1-x_2}{1-x_1}\ln(\frac{x_1-x_2}{1-x_2})\right] ~\nonumber \\
&&+\frac{i}{16\pi^2}x_2\left[L+1-\ln(\frac{m_{\pi}^{2}}{\mu^2})+\frac{x_1-x_2}{x_2}\ln(\frac{x_1-x_2}{x_1})\right]
-\frac{1}{16\pi}(x_1-x_2) ~\nonumber \\
&=& \frac{i}{16\pi^2}\left[L-\ln(\frac{m_{\pi}^{2}}{m_{\pi_1}^2})\right](a+ab)+\frac{i}{16\pi^2}\left[L-\ln(\frac{m_{\eta}^{2}}{m_{\pi_1}^2})\right](b+ab) ~\nonumber \\
&&+\frac{i}{32\pi^2}(a^2+b^2)-\frac{1}{16\pi}(1-a-b-2ab).
\end{eqnarray}

The chiral correction from the $\eta \pi$ loop diagram reads
\begin{eqnarray}
\Sigma_{T,IR}^{\pi\eta}(m_{\pi_{1}}^{2}) &=&
\frac{g_{\pi\eta}^{2}m_{\pi}^{4}}{128\pi^{2}m_{\pi_{1}}^{2}}\left[1-2\ln(\frac{m_{\pi}^{2}}{m_{\pi_{1}}^{2}})\right]
+\frac{g_{\pi\eta}^{2}m_{\eta}^{4}}{128\pi^{2}m_{\pi_{1}}^{2}}\left[1-2\ln(\frac{m_{\eta}^{2}}{m_{\pi_{1}}^{2}})\right]~\nonumber \\
&&-i g_{\pi\eta}^{2}\left(\frac{m_{\pi_1}^2-3m_{\pi}^2-3m_{\eta}^2}{192\pi}+\frac{m_{\pi}^4+m_{\eta}^4}{64\pi m_{\pi_1}^2}\right).
\end{eqnarray}

\subsection{$\eta(\eta')$-$\pi_1$ loop}

The $\eta'$ meson mass is dominated by the axial anomaly, which
remains large in the chiral limit. The propagators in the
$\eta'$-$\pi_1$ loop do not produce a 'soft pole' contribution. In
other words, the loop integral does not contain the 'infrared
singular part'.

Now we consider the $\pi_1 \eta$ loop diagram with $M_{\pi_1}^2\sim
M^2$ and $ m^2\ll M^2 $, which is similar to the nucleon self energy
diagram. We can use the standard IR method in Ref.
\cite{Becher:1999he} to obtain the 'infrared singular part'. First we
define the dimensionless variables
\begin{equation}
\Omega=\frac{p^2-m_\eta^2-m_{\pi_1}^2}{2 m_\eta m_{\pi_1}},~~~~
\alpha =\frac{m_\eta}{m_{\pi_1}}.
\end{equation}
The corresponding scalar loop integral is
\begin{eqnarray}
I_{\pi_1 \eta}(p^2)&=&\mu^{4-d}\int
\frac{d^dl}{(2\pi)^d}\frac{1}{[l^2-m_{\eta}^2+i\epsilon][(p-l)^2-m_{\pi_1}^2+i\epsilon]}~\nonumber \\
&=& \mu^{4-d}\Gamma \left(2-\frac{d}{2}\right)\frac{i
m_{\pi_1}^{d-4}}{(4\pi)^{\frac{d}{2}}}\int_{0}^{1}dx
(\Delta)^{\frac{d}{2}-2},
\end{eqnarray}
where
\begin{equation}
\Delta = x^{2}-2\alpha\Omega x(1-x)+\alpha^{2}(1-x)^{2}-i\epsilon.
\end{equation}

Within the IR scheme, the 'infrared singular part' of $I_{\pi_1 \eta}$
reads
\begin{eqnarray}
I_{\pi_1 \eta}^{IR}&=&\mu^{4-d}\Gamma \left(2-\frac{d}{2}\right)\frac{i m_{\pi_1}^{d-4}}{(4\pi)^{\frac{d}{2}}}\int_{0}^{\infty}dx (\Delta)^{\frac{d}{2}-2}~\nonumber\\
  &=& \frac{i(p^{2}-m_{\pi_1}^{2}+m_\eta^{2})}{32\pi^2 p^{2}}L + I'(p^{2})
\end{eqnarray}
with
\begin{eqnarray}
I'(p^{2})=
\frac{i}{16\pi^2}\frac{\alpha(\alpha+\Omega)}{1+2\alpha\Omega
+\alpha^2}\left(1-2\ln\alpha\right)-\frac{i}{8\pi^2}\frac{\alpha
  \sqrt{1-\Omega^2}} {1+2\alpha\Omega + \alpha^2}
\arccos\left(-\frac{\alpha+\Omega}{\sqrt{1+2\alpha\Omega+\alpha^2}}\right),
\end{eqnarray}
and the regularization scale $\mu =m_{\pi_1}$. The chiral correction
from the $\pi_1 \eta$ loop diagram reads
\begin{eqnarray}
\Sigma_{T,IR}^{\pi_{1}\eta}(m_{\pi_{1}}^{2})=-g_{\pi_{1}\eta}^{2}\left[\frac{m_{\pi_{1}}m_{\eta}^{3}}{24\pi}
+\frac{m_{\eta}^4}{32\pi^2}\ln(\frac{m_{\eta}^{2}}{m_{\pi_{1}}^{2}})\right]+{\cal
O}(m_{\eta}^5).
\end{eqnarray}

\subsection{Tadpole diagrams}

\begin{figure}
\centering
\includegraphics[width=0.6\linewidth]{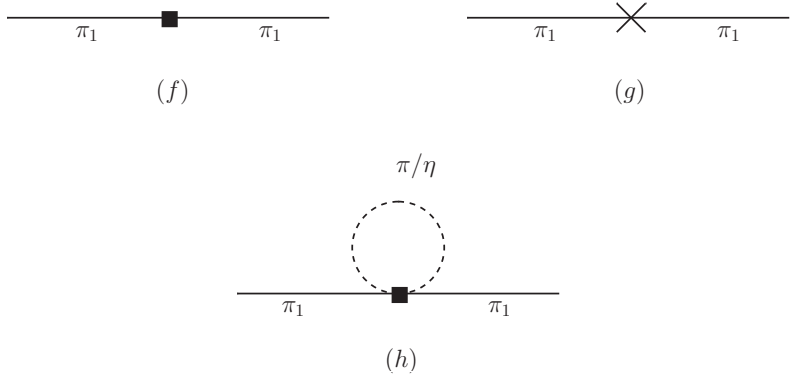}
\caption{The tadpole diagram of the $\pi_1(1600)$ self energy. The
$\cal O$$(m_{\pi}^2)$ and $\cal O$$(m_{\pi}^4)$ LECs also contribute
to the self energy, which are labeled by square and cross
respectively.} \label{counterterms}
\end{figure}

The chiral corrections from the tadpole diagrams in Fig.
\ref{counterterms} are
\begin{eqnarray}
\Sigma_{T,IR}^{\pi,tadpole}(m_{\pi_{1}}^{2})=(d_1+\frac{d_2}{4})\frac{3m_{\pi}^{4}}{16\pi^2}\ln(\frac{m_{\pi}^2}{m_{\pi_1}^2})
-\frac{3}{128\pi^2}d_2 m_{\pi}^4,
\end{eqnarray}
\begin{eqnarray}
\Sigma_{T,IR}^{\eta,tadpole}(m_{\pi_{1}}^{2})=(d_1^{*}+\frac{d_2^{*}}{4})\frac{m_{\eta}^{4}}{16\pi^2}\ln(\frac{m_{\eta}^2}{m_{\pi_1}^2})
-\frac{1}{128\pi^2}d_2^{*} m_{\eta}^4,
\end{eqnarray}
where we have redefined the low energy constants
\begin{eqnarray}
d_1=c_1+c_2+c_6,~~~~d_2=c_3,\nonumber \\
d_1^*=c_1^*+c_2^*+c_6^*,~~~~d_2^*=c_3^*.
\end{eqnarray}

All the divergence can be absorbed by the counter terms in Eq.
\ref{counter}, which also contribute to $m_{\pi_1}$
\begin{eqnarray}
\Sigma_{\pi_1(1600)}^{tree} = e_{1}(m_\pi^2+m_\eta^2)+
e_{2}(m_\pi^2+m_\eta^2)^2.
\end{eqnarray}

Finally we obtain the chiral corrections to the $\pi_1(1600)$ mass
up to the one loop order, which is the main result of this work
\begin{eqnarray}
\label{final} \Delta M_{\pi_1(1600)}^{1-loop} &=&
\Sigma_{T,IR}^{\rho\pi}(m_{\pi_{1}}^{2})+\Sigma_{T,IR}^{\pi_{1}\eta}(m_{\pi_{1}}^{2})
+\Sigma_{T,IR}^{\pi_{1}\eta'}(m_{\pi_{1}}^{2})+\Sigma_{T,IR}^{\pi\eta}(m_{\pi_{1}}^{2})
+\Sigma_{T,IR}^{\pi\eta'}(m_{\pi_{1}}^{2})~\nonumber \\
&&+\Sigma_{T,IR}^{b_{1}\pi}(m_{\pi_{1}}^{2})+\Sigma_{T,IR}^{f_{1}\pi}(m_{\pi_{1}}^{2})
+\Sigma_{T,IR}^{\pi,tadpole}(m_{\pi_{1}}^{2})+\Sigma_{T,IR}^{\eta,tadpole}(m_{\pi_{1}}^{2})+\Sigma_{\pi_1(1600)}^{tree}.
\end{eqnarray}
One may note that we treat the intermediate states as stable particles in our above calculation,
however, the widths of $\rho,~b_1,~f_1$ are not small. The contributions from the widths of
the intermediate states to the non-analytic chiral corrections to the $\pi_1(1600)$ mass
are summarized in Appendix \ref{appendix}.

%%%%%%%%%%%%%%%%%%%%%%%%%%%%%%%%%%%%%%%%%%%%%%%%%
\section{Results and discussions}\label{sec4}
%%%%%%%%%%%%%%%%%%%%%%%%%%%%%%%%%%%%%%%%%%%%%%%%%

We need to deal with the numerous effective coupling constants
before the numerical analysis. Actually the experimental information
on the $\pi_1(1600)$ decays is not rich. From the current
experimental data of the $\pi_1(1600)$ decays, we can make a very
rough estimate of the values of
$g_{\rho\pi},~g_{\eta\pi},~g_{\eta'\pi},~g_{f_{1}\pi}$~and~$g_{b_{1}\pi}$.
The others still remain unknown.

A partial wave analysis in Ref. \cite{Zaitsev:2000rc} gives the
branching ratio
\begin{eqnarray}
Br(\pi_1\rightarrow b_1\pi):Br(\pi_1\rightarrow
\rho\pi):Br(\pi_1\rightarrow \eta'\pi)=1:(1.5\pm0.5):(1.0\pm0.3).
\end{eqnarray}
The analysis based on the VES experiment leads to
\cite{Amelin:2005ry}
\begin{eqnarray}
Br(\pi_1\rightarrow b_1\pi):Br(\pi_1\rightarrow
\rho\pi):Br(\pi_1\rightarrow \eta'\pi):Br(\pi_1\rightarrow
f_1\pi)=(1.0\pm0.3):<0.3:1:(1.1\pm0.3).
\end{eqnarray}
The E852 collaboration reported \cite{Kuhn:2004en}
\begin{eqnarray}
\frac{Br(\pi_1\rightarrow f_1\pi)}{Br(\pi_1\rightarrow
\eta'\pi)}=3.80\pm0.78.
\end{eqnarray}
In order to make a very rough estimate of these coupling constants,
we combine the above measurements and set the branching ratio to be
\begin{eqnarray}
\label{ratio} Br(\pi_1\rightarrow b_1\pi):Br(\pi_1\rightarrow
\rho\pi):Br(\pi_1\rightarrow \eta'\pi):Br(\pi_1\rightarrow
f_1\pi):Br(\pi_1\rightarrow \eta\pi)= 1:2:1:1:1.
\end{eqnarray}

From Eqs. (\ref{Lagrangian1})- (\ref{Lagrangian2}), the partial
decay width of the $\pi_1(1600)$ reads
\begin{eqnarray}
\Gamma(\pi_{1}\rightarrow\rho\pi)=2\times\frac{g_{\rho\pi}^{2}}{12\pi}|\vec{\text{p}}_{\pi}|^{3},
\end{eqnarray}
\begin{eqnarray}
\Gamma(\pi_{1}\rightarrow\eta\pi)=\frac{g_{\eta\pi}^{2}}{24\pi}\frac{|\vec{\text{p}}_{\pi}|^{3}}{m_{\pi_{1}}^{2}},
\end{eqnarray}
\begin{eqnarray}
\Gamma(\pi_{1}\rightarrow\eta'\pi)=\frac{g_{\eta'\pi}^{2}}{24\pi}\frac{|\vec{\text{p}}_{\pi}|^{3}}{m_{\pi_{1}}^{2}},
\end{eqnarray}
\begin{eqnarray}
\Gamma(\pi_{1}\rightarrow
f_{1}\pi)=\frac{g_{f_{1}\pi}^{2}}{24\pi}\frac{|\vec{\text{p}}_{\pi}|}{m_{\pi_{1}}^{2}}(3+\frac{|\vec{\text{p}}_{\pi}|^{2}}{m_{f_{1}}^{2}}),
\end{eqnarray}
\begin{eqnarray}
\Gamma(\pi_{1}\rightarrow
b_{1}\pi)=2\times\frac{g_{b_{1}\pi}^{2}}{24\pi}\frac{|\vec{\text{p}}_{\pi}|}{m_{\pi_{1}}^{2}}(3+\frac{|\vec{\text{p}}_{\pi}|^{2}}{m_{b_{1}}^{2}}),
\end{eqnarray}
where $\vec{\text{p}}_{\pi}$ is the pion decay momentum.

With the total decay width of $\pi_1(1600)$ around 300 MeV as input
\cite{Agashe:2014kda}, we get
\begin{eqnarray}
|g_{\rho\pi}| \simeq
2.7~\mbox{GeV}^{-1},~~~|g_{\eta\pi}| \simeq
5.1,~~~|g_{\eta'\pi}| \simeq 8.1,
~~~|g_{f_1\pi}| \simeq 3.3~\mbox{GeV},~~~|g_{b_1\pi}| \simeq
2.2~\mbox{GeV}.
\end{eqnarray}
For the $\pi_1\pi_1\eta$ coupling constant, we use
$g_{\pi_1\eta}\sim \frac{1}{1.6 F_\eta }~\mbox{GeV}^{-1}\sim
5.3~\mbox{GeV}^{-1}$ where the $F_\eta\approx$~0.1 GeV is the decay
constant of $\eta$. This ad hoc value was estimated with the very naive
dimensional argument, which might be too large.

\begin{figure}
\centering
\includegraphics[width=0.4\linewidth]{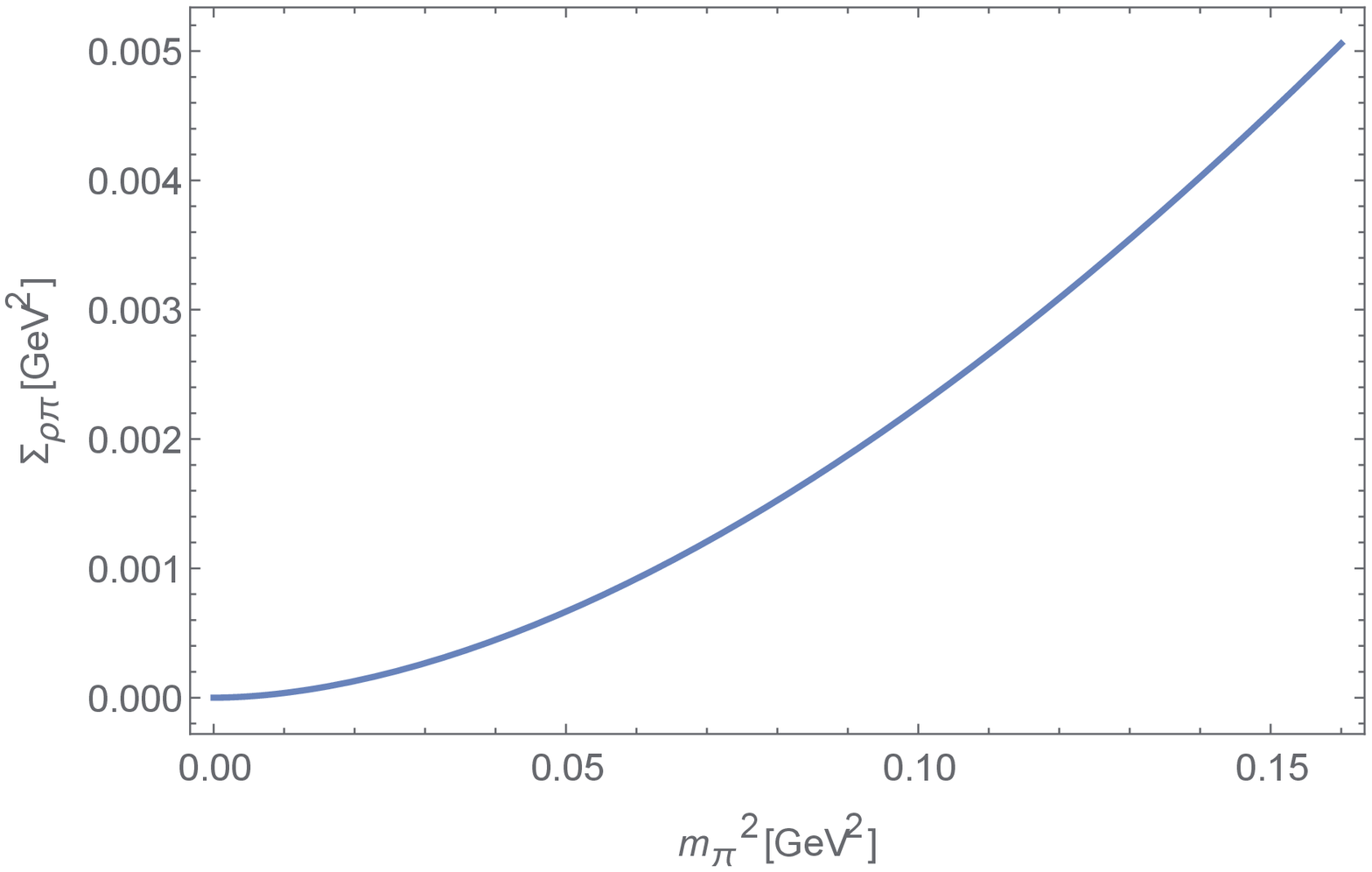}
\includegraphics[width=0.4\linewidth]{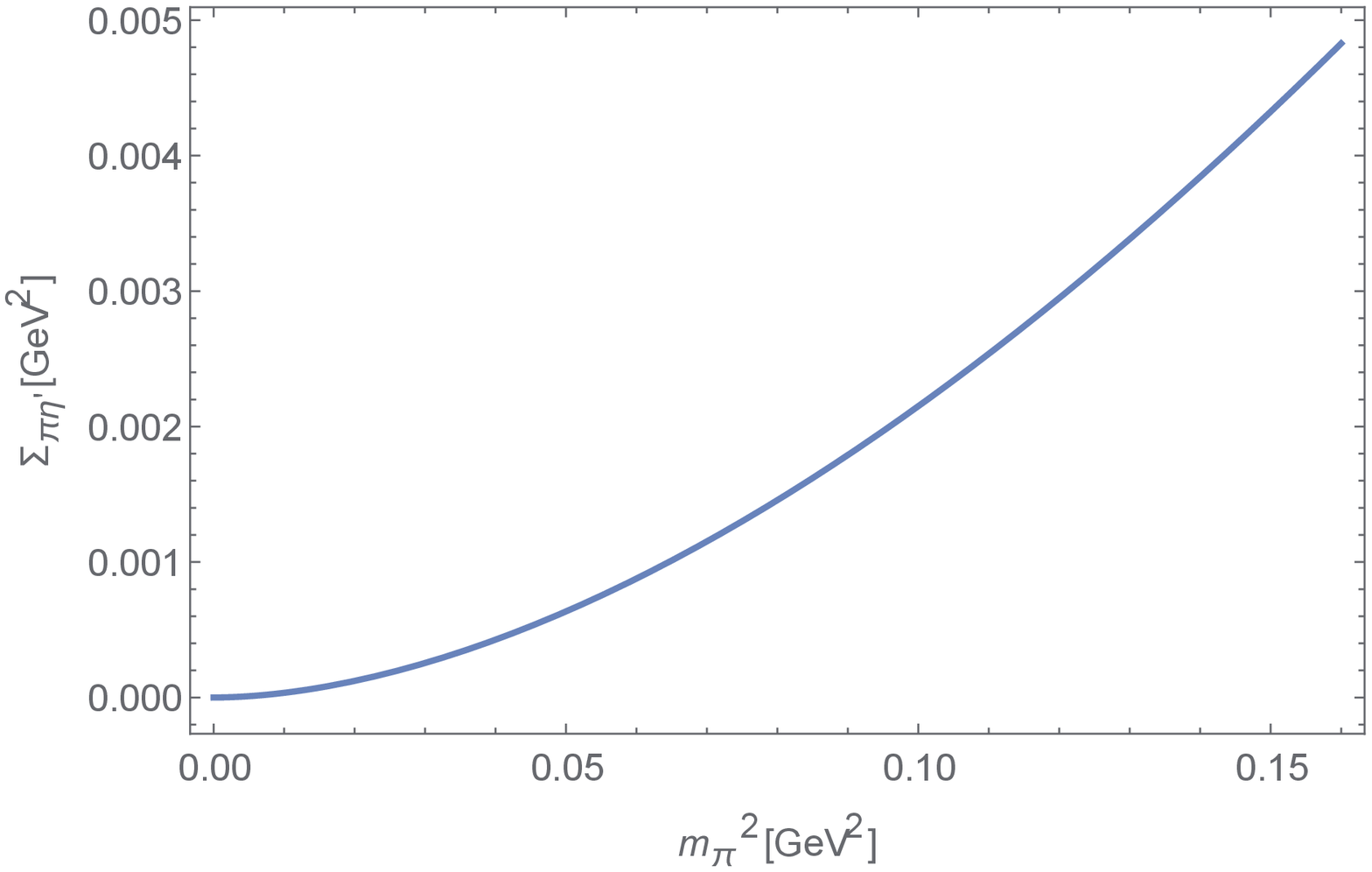}
\includegraphics[width=0.4\linewidth]{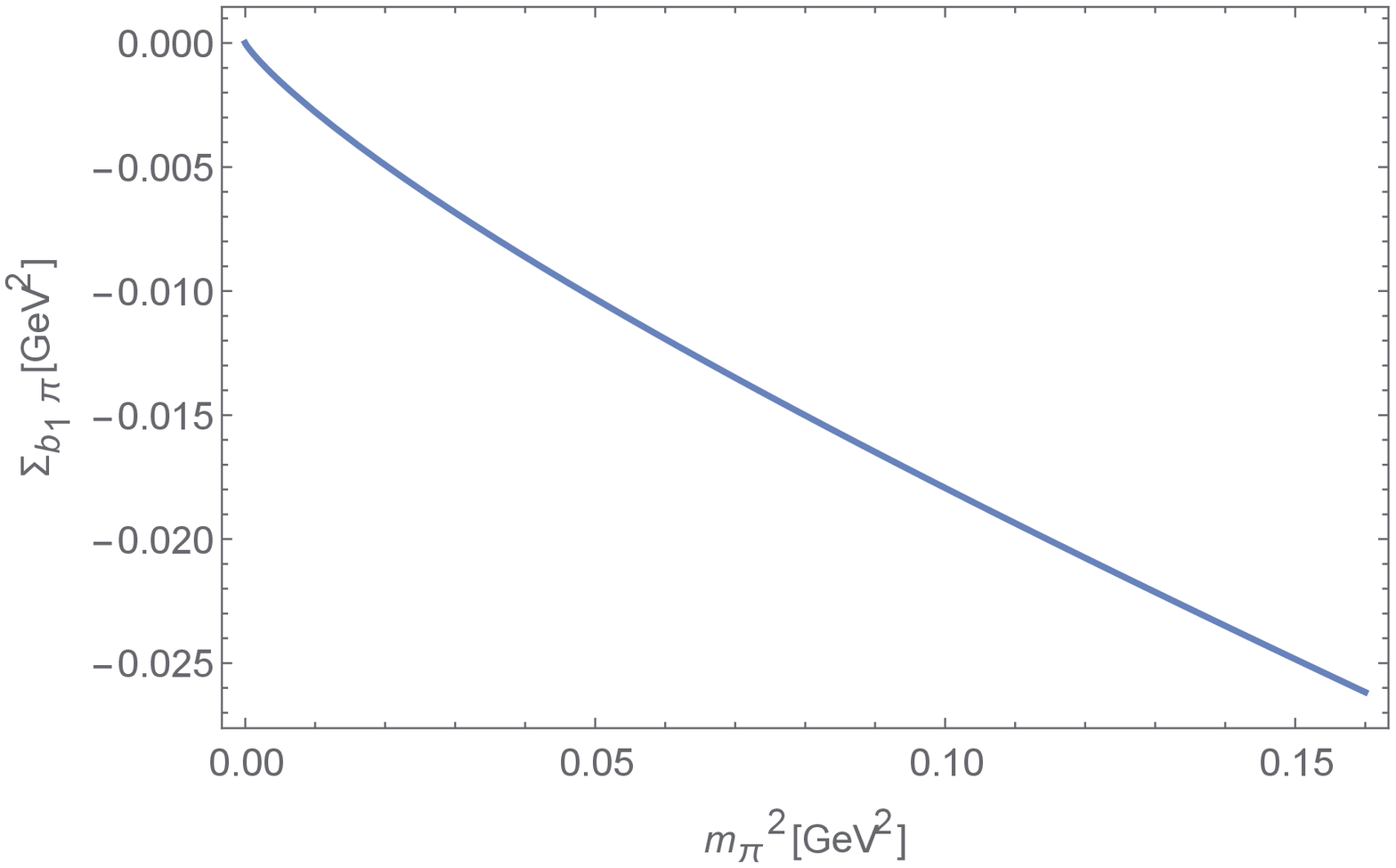}
\includegraphics[width=0.4\linewidth]{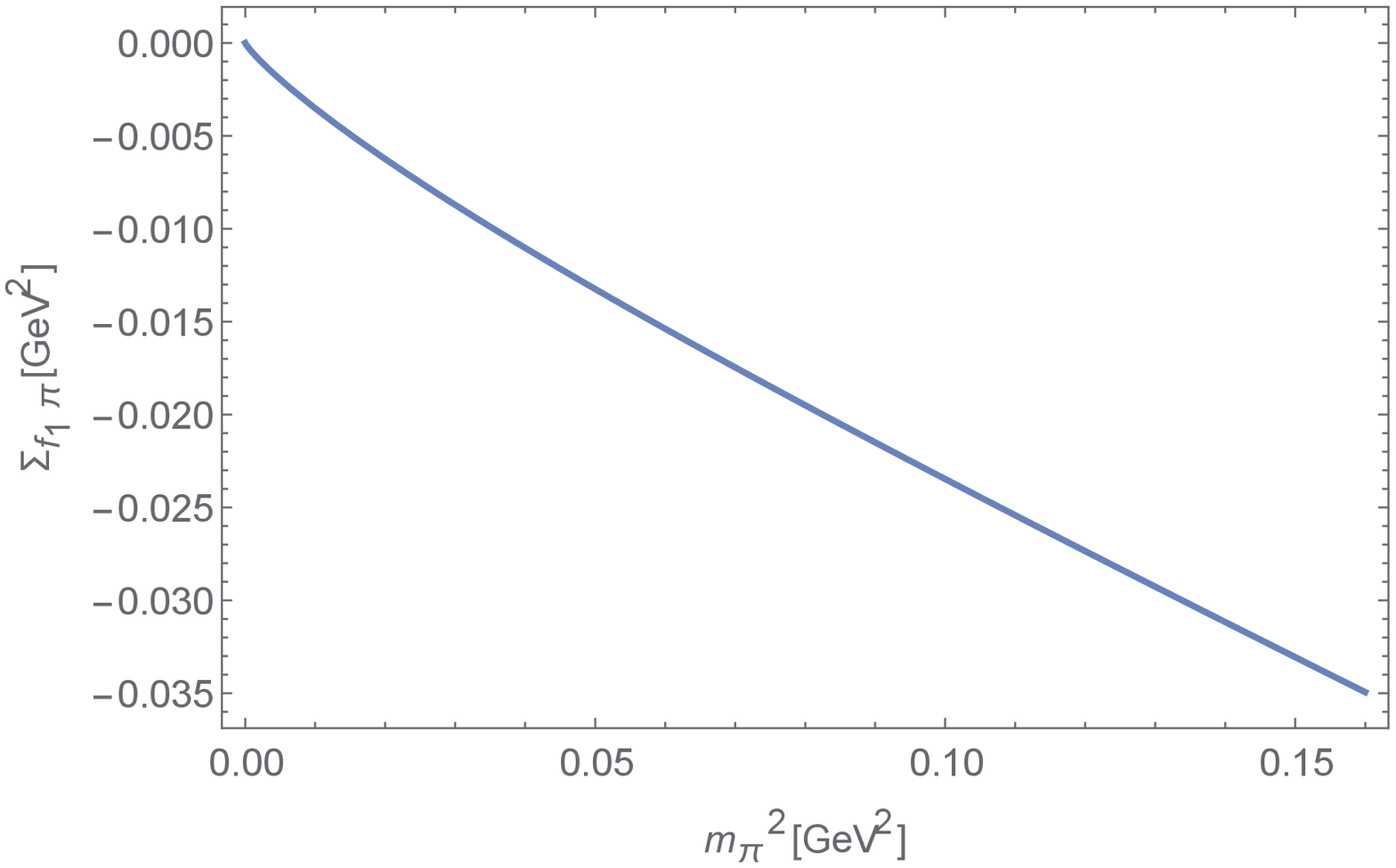}
\caption{The pion mass dependence of the chiral corrections to the
$\pi_1(1600)$ mass from the $\rho\pi, \eta'\pi, b_1\pi, f_1\pi$
loops, where the top-left, top-right, bottom-left and bottom-right subfigures correspond to the $\rho\pi, \eta'\pi, b_1\pi, f_1\pi$ contributions respectively.} \label{curves}
\end{figure}

\begin{figure}
\centering
\includegraphics[width=0.4\linewidth]{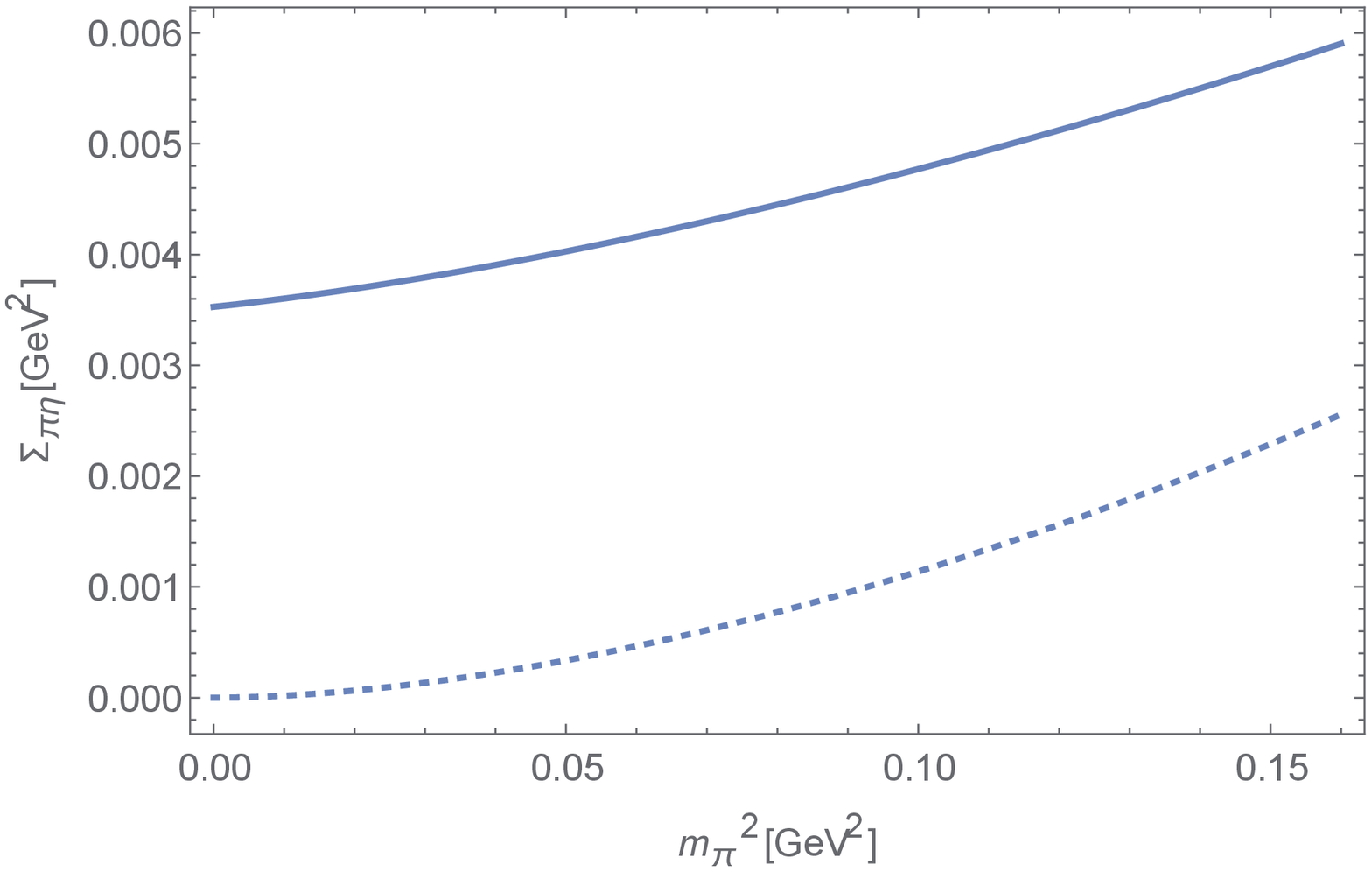}
\includegraphics[width=0.4\linewidth]{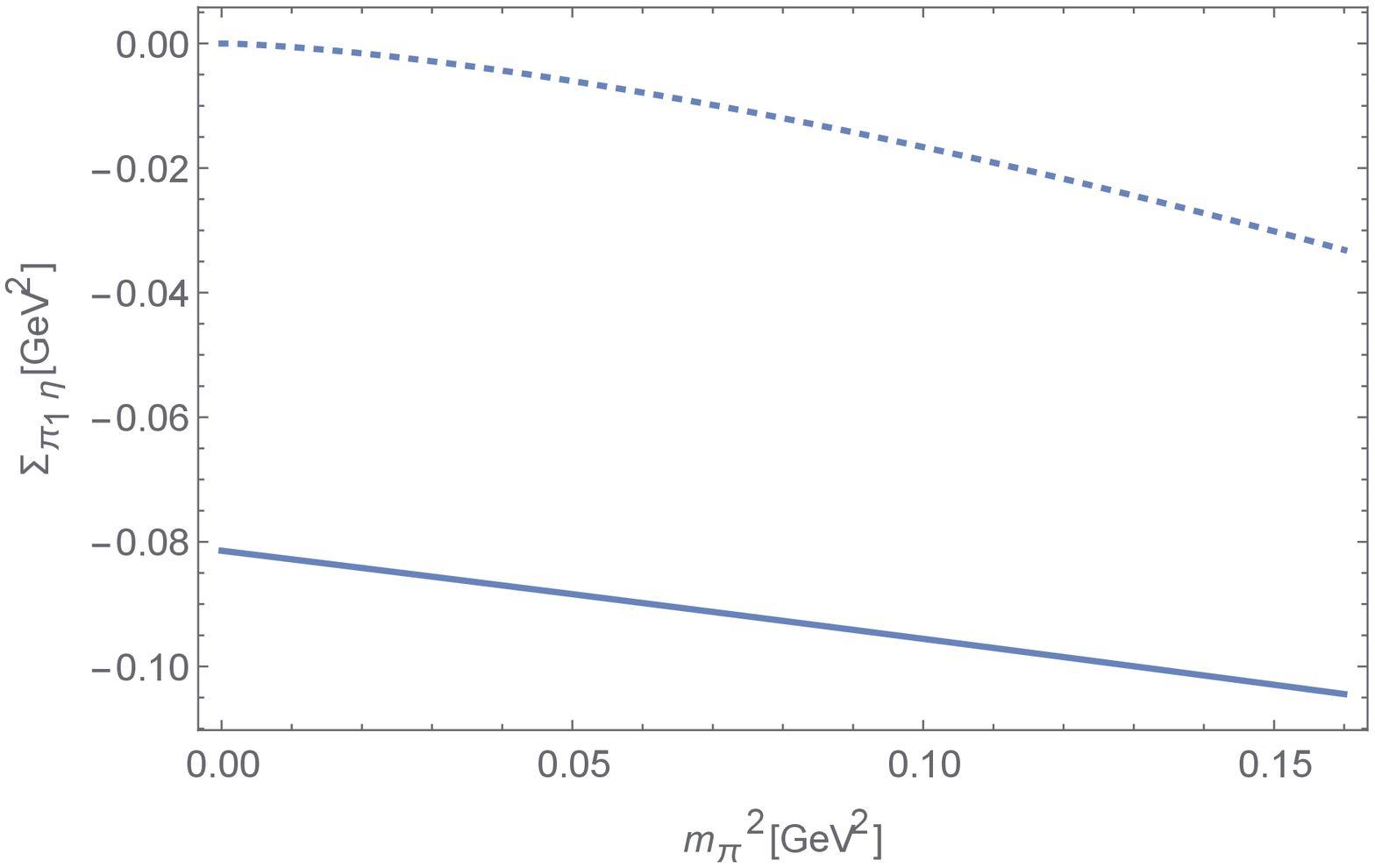}
\caption{The pion mass dependence of the chiral corrections to the
$\pi_1(1600)$ mass from the $\eta\pi$ and $\pi_1\eta$ loops. The
solid and dotted lines correspond to the $SU_F(2)$ and $SU_F(3)$
cases respectively.} \label{curves1}
\end{figure}

From the tree-level Lagrangian of chiral perturbation theory,
\begin{eqnarray}
M_{\pi}^2=2B_0m,~~~ M_{\eta}^2=\frac{2}{3}B_0(m+2m_s).
\end{eqnarray}
We consider two cases in the numerical analysis. Case 1 corresponds
to the $SU_F(3)$ chiral limit where $M_{\pi}^2=M_{\eta}^2 \to 0$
when $m_s=m$ approaches zero simultaneously.

Since the strange quark is sometimes treated as a heavy degree of
freedom in the lattice QCD simulation, we also consider Case 2,
which corresponds to the $SU_F(2)$ chiral limit. Now we fix the
strange quark mass and let the up and down quark mass approach zero.
In the $SU_F(2)$ chiral limit, the $\eta$ meson mass remains finite.
We have
\begin{eqnarray}
M_{\eta}^2=\frac{4}{3}B_0m_s+\frac{1}{3}M_{\pi}^2.
\end{eqnarray}

We collect the variation of the chiral corrections to the
$\pi_1(1600)$ mass from different loop diagrams with the pion mass
in Figs. (\ref{curves})-(\ref{curves1}). The most important chiral
correction to the $\pi_1(1600)$ mass comes from the $\pi_1 \eta$
loop. The chiral corrections from the $\pi \rho,~\pi \eta$ and $\pi
\eta'$ loops are positive and increase with $m_{\pi}$ while
the corrections from the
$\eta \pi_1,~\pi b_1$ and $\pi f_1$ loops are negative. On
the other hand, the chiral corrections from the $\eta \pi_1,~\pi
b_1$ and $\pi f_1$ loops are very sensitive to the pion mass.

The coupling constants $d_i ~(i=1,2)$, $d_j^* ~(j=1,2)$
contribute to the tadpole diagram while $e_k ~(k=1,2)$ are the low
energy constants. They are unknown at present. Although this kind of
contribution may be significant, we do not present their variations
with the pion mass because of too many unknown coupling constants.

According to PDG \cite{Amsler:2008zzb}, the $\pi_1(1600)$ was
observed in the $b_1\pi$, $\eta' \pi$ and $f_1\pi$ modes. The
Compass collaboration reported the $\pi_1(1600)$ in the $\rho\pi$
mode \cite{Alekseev:2009aa}. On the other hand, the $\pi_1(1400)$
was observed in the $\eta\pi$ mode. Both the $\pi_1(1600)$ and
$\pi_1(1400)$ signals are very broad with a decay width of $241\pm
40$ MeV and $330\pm 35$ MeV respectively \cite{Amsler:2008zzb}.
These two signals overlap with each other. In this work, we have
taken into account all the above possible decay modes and calculated
the one-loop chiral corrections to the $\pi_1(1600)$ mass. We have
employed two different methods to deal with the loop integrals and
derived all the infrared singular chiral corrections explicitly.

From the available experimental measurement of the partial decay
width of the $\pi_1(1600)$ meson, we extract the coupling constants.
We investigate the variation of the different chiral corrections
with the pion mass under two schemes. The present calculation is
applicable to all possible interpretations of the $\pi_1$ mesons
since our analysis does not rest on the inner structure of the
$\pi_1$ mesons. Hopefully, the explicit non-analytical chiral
structures will be helpful to the chiral extrapolation of the
lattice data from the dynamical lattice QCD simulation of either the
exotic light hybrid meson or tetraquark state.

%*******************************************
\section*{Acknowledgements}
%*******************************************

This project is supported by National Natural Science Foundation of
China under Grants No. 11222547, No. 11175073, No. 11575008 and 973
program. XL is also supported by the National Youth Top-notch Talent
Support Program ("Thousands-of-Talents Scheme").

%%%%%%%%%%%%%%%%%%%%%%%%%%%%%%%%%%%%%%%%%%%%%%%%%%%%%%%%%%%%%%%%%%%%%%%%%%%%%%%%
\appendix
%*******************************************
\section{Contributions generated by the finite widths of the intermediate states}\label{appendix}
%*******************************************

In this Appendix we deal with the scalar loop integrals when the
intermediate states have a finite decay width $\Gamma$.
\begin{eqnarray}
I_{\pi X}(p^2)&=&\mu^{4-d}\int
\frac{d^dl}{(2\pi)^d}\frac{1}{[l^2-m_\pi^2+i\epsilon][(p-l)^2-M^2+iM\Gamma]}~\nonumber \\
&=& \mu^{4-d}\Gamma \left(2-\frac{d}{2}\right)\frac{i
M^{d-4}}{(4\pi)^{\frac{d}{2}}}\int_{0}^{1}dx
(\Delta)^{\frac{d}{2}-2},
\end{eqnarray}
with
\begin{eqnarray}
\Delta &=& b x^2-(a+b-1+\frac{i\Gamma}{M})x+a~\nonumber\\
&=&b(x-x_1)(x-x_2),~\nonumber\\
a&=& \frac{m_\pi^{2}}{M^{2}}~,~~~~~~~~ b = \frac{p^{2}}{M^{2}},
\end{eqnarray}
where the $X$ represents $\rho,~b_1,~f_1$, the $M$ and $\Gamma$ are
the corresponding mass and width, and
\begin{eqnarray}
x_{1,2} &=&  \frac{a+b-1+\frac{i\Gamma}{M} }{2 b} \left( 1 \pm
         \sqrt{1 - \frac{4 a b}{(a+b-1+\frac{i\Gamma}{M})^2} }  \  \right)~.
\end{eqnarray}
We expand $x_{1,2}$ in terms of $a$
\begin{eqnarray}
x_1&=&\frac{b-1+\frac{i\Gamma}{M}}{b}-\frac{a(1-\frac{i\Gamma}{M})}{b(b-1+\frac{i\Gamma}{M})}
-\frac{a^2(1-\frac{i\Gamma}{M})}{(b-1+\frac{i\Gamma}{M})^3}+{\cal O}(a^3),~\nonumber \\
x_2&=&\frac{a}{b-1+\frac{i\Gamma}{M}}+\frac{a^2(1-\frac{i\Gamma}{M})}{(b-1+\frac{i\Gamma}{M})^3}+{\cal O}(a^3).
\end{eqnarray}
In our case, the $\Gamma_{X}\sim m_\pi$. We treat the
$(\frac{\Gamma}{M})^2$ as ${\cal O}(a)$ and get
\begin{eqnarray}
x_1&=&\frac{b-1}{b}-\frac{a(b-1-\frac{\Gamma^2}{M^2})}{b[(b-1)^2+\frac{\Gamma^2}{M^2}]}
-\frac{a^2(b-1)^3}{[(b-1)^2+\frac{\Gamma^2}{M^2}]^3}
+i \frac{\Gamma}{M}\left[\frac{1}{b}+\frac{a}{(b-1)^2+\frac{\Gamma^2}{M^2}}\right]+\ldots,~\nonumber \\
x_2&=&\frac{a(b-1)}{(b-1)^2+\frac{\Gamma^2}{M^2}}+\frac{a^2(b-1)^3}{[(b-1)^2+\frac{\Gamma^2}{M^2}]^3} -i\frac{\Gamma}{M}\frac{a}{(b-1)^2+\frac{\Gamma^2}{M^2}}+\ldots.
\end{eqnarray}
The original integral can be re-expressed as
\begin{eqnarray}
I_{\pi X}(p^2) &=& \mu^{4-d}\Gamma \left(2-\frac{d}{2}\right)\frac{i
M^{d-4}}{(4\pi)^{\frac{d}{2}}}\int_{0}^{1}dx
[b(x-x_1)(x-x_2)]^{\frac{d}{2}-2}.
\end{eqnarray}
Now $x_1$, $x_2$ are complex while the integration variable $x$ is
real, which renders the evaluation of the integral straightforward.
We have
\begin{eqnarray}
I_{\pi X}(p^2) &=& \frac{i}{16\pi^2}\left[L-\ln(\frac{M^2}{\mu^2})-1-\int_0^1 dx\ln[b(x-x_1)(x-x_2)]\right]\nonumber\\
&=&\frac{i}{16\pi^2}(1-x_2)\left[L-\ln(\frac{M^2}{\mu^2})\right]+\frac{i}{16\pi^2}x_2\left[L-\ln(\frac{m_\pi^{2}}{\mu^2})\right]~\nonumber\\
&&+\frac{i}{16\pi^2}\left[1-(1-x_2)\ln(1-\frac{i\Gamma}{M})-(x_1-x_2)\ln(\frac{-x_1}{1-x_1})\right].
\end{eqnarray}
After extracting the non-analytic chiral corrections from the above
expression, we get
\begin{eqnarray}
I_{\pi X}^{NA}(p^2) &=& -\frac{i}{16\pi^2}x_2\ln(\frac{m_\pi^{2}}{\mu^2})~\nonumber \\
&=&-\frac{i}{16\pi^2}\left[\frac{a(b-1)}{(b-1)^2+\frac{\Gamma^2}{M^2}}+\frac{a^2(b-1)^3}{[(b-1)^2+\frac{\Gamma^2}{M^2}]^3} -i\frac{\Gamma}{M}\frac{a}{(b-1)^2+\frac{\Gamma^2}{M^2}}\right]\ln(\frac{m_\pi^{2}}{\mu^2}).
\end{eqnarray}
It's interesting to note that the above expression contains a
non-analytic chiral correction to the imaginary part, which is
proportional to $\frac{\Gamma}{M}$ and vanishes when $\Gamma\to 0$.
In comparison, when we treat the intermediate states as stable
particles, the imaginary parts of the chiral corrections to the
self-energy of the $\pi_1(1600)$ are analytic in the pseudo scalar
meson mass. In the limit of $\Gamma = 0$, we recover the results in
the previous sections in the text.

For the $\rho\pi,~b_1\pi,~f_1\pi$ loops, we collect the non-analytic
chiral corrections to the mass of the $\pi_1(1600)$ up to ${\cal
O}(m_{\pi}^4)$,
\begin{eqnarray}
\Sigma_{T,NA}^{\rho\pi}(m_{\pi_{1}}^{2})&=&-\frac{g_{\rho\pi}^{2}m_{\pi}^2}{48\pi^2}\ln(\frac{m_{\pi}^2}{m_{\pi_1}^2})\left\{\frac{m_{\rho}^2 \Gamma_{\rho}^2(m_{\pi_1}^2-m_{\rho}^2)}{(m_{\pi_1}^2-m_{\rho}^2)^2+m_{\rho}^2 \Gamma_{\rho}^2}\right.~\nonumber \\&&
+\left.\frac{m_{\pi}^2(3m_{\pi_1}^4-2m_{\pi_1}^2m_{\rho}^2+m_{\rho}^2 \Gamma_{\rho}^2-m_{\rho}^4)}{(m_{\pi_1}^2-m_{\rho}^2)^2+m_{\rho}^2 \Gamma_{\rho}^2}
-\frac{m_{\pi}^2 m_{\rho}^2(m_{\pi_1}^2-m_{\rho}^2)^5}{[(m_{\pi_1}^2-m_{\rho}^2)^2+m_{\rho}^2 \Gamma_{\rho}^2]^3}\right\},\\
\Sigma_{T,NA}^{b_1\pi}(m_{\pi_{1}}^{2})&=&\frac{g_{b_1\pi}^{2} m_{\pi}^2}{96\pi^2 m_{\pi_1}^2}\ln(\frac{m_{\pi}^2}{m_{\pi_1}^2})
\left\{\frac{(m_{\pi_1}^2-m_{b_1}^2)(12m_{\pi_1}^2-\Gamma_{b_1}^2)}{(m_{\pi_1}^2-m_{b_1}^2)^2+m_{b_1}^2 \Gamma_{b_1}^2}\right.~\nonumber \\&&
-\frac{m_{\pi}^2(3m_{\pi_1}^4-2m_{\pi_1}^2m_{b_1}^2+m_{b_1}^2 \Gamma_{b_1}^2-m_{b_1}^4)}{m_{b_1}^2 [(m_{\pi_1}^2-m_{b_1}^2)^2+m_{b_1}^2 \Gamma_{b_1}^2]}~\nonumber \\&&
+\left.\frac{m_{\pi}^2(m_{\pi_1}^2-m_{b_1}^2)^3(m_{\pi_1}^4+10m_{\pi_1}^2 m_{b_1}^2+m_{b_1}^4)}{[(m_{\pi_1}^2-m_{b_1}^2)^2+m_{b_1}^2 \Gamma_{b_1}^2]^3}\right\},\\
\Sigma_{T,NA}^{f_1\pi}(m_{\pi_{1}}^{2})&=&\frac{g_{f_1\pi}^{2} m_{\pi}^2}{192\pi^2 m_{\pi_1}^2}\ln(\frac{m_{\pi}^2}{m_{\pi_1}^2})
\left\{\frac{(m_{\pi_1}^2-m_{f_1}^2)(12m_{\pi_1}^2-\Gamma_{f_1}^2)}{(m_{\pi_1}^2-m_{f_1}^2)^2+m_{f_1}^2 \Gamma_{f_1}^2}\right.~\nonumber \\&&
-\frac{m_{\pi}^2(3m_{\pi_1}^4-2m_{\pi_1}^2m_{f_1}^2+m_{f_1}^2 \Gamma_{f_1}^2-m_{f_1}^4)}{m_{f_1}^2 [(m_{\pi_1}^2-m_{f_1}^2)^2+m_{f_1}^2 \Gamma_{f_1}^2]}~\nonumber \\&&
+\left.\frac{m_{\pi}^2(m_{\pi_1}^2-m_{f_1}^2)^3(m_{\pi_1}^4+10m_{\pi_1}^2 m_{f_1}^2+m_{f_1}^4)}{[(m_{\pi_1}^2-m_{f_1}^2)^2+m_{f_1}^2 \Gamma_{f_1}^2]^3}\right\}.
\end{eqnarray}

%%%%%%%%%%%%%%%%%%%%%%%%%%%%%%%%%%%%%%%%%%%%%%%%%%%%%%%%%%%%%%%%%%%%%%%%%%%%%%%%

\end{document}